\documentclass[11pt]{article}
\usepackage {amsfonts}
\usepackage {graphicx}
\usepackage {epsfig}
\usepackage{amssymb}
\usepackage{amsmath}
\usepackage {longtable}
\usepackage[english]{babel}
\begin{document}

\title{Spin rotation of polarized beams in high energy storage rings}

\author{V.G. Baryshevsky \\ Research Institute for Nuclear Problems, Belarusian State
University,\\ 11 Bobruyskaya Str., Minsk 220050, Belarus,
\\ e-mail: bar@inp.minsk.by}

\maketitle

\begin{center}
\begin{abstract}
The equations for spin evolution of a particle in a storage ring
are obtained considering contributions from the tensor electric
and magnetic polarizabilities of the particle along with the
contributions from spin rotation and birefringence effect in
polarized matter of an internal target.
Study of the spin rotation and birefringence effects for a
particle in a high energy storage ring provides for measurement
both the spin-dependent real part of the coherent elastic
zero-angle scattering amplitude and tensor electric (magnetic)
polarizabilities.

\end{abstract}
\end{center}

\section{INTRODUCTION}

Investigation of spin-dependent interactions of elementary
particles at high energies is a very important part of programs
for scientific research at storage rings \cite{Lehar,STORI-FAIR}.
Such studies are being carried with the use of polarized beams and
polarized targets.
Dependence of scattering cross-sections on the particle spin is
the subject of much studies.
The experiments for measuring the spin-dependent part of the
forward scattering amplitude are in preparation now
\cite{STORI-FAIR}.

 It should be mentioned that it is well known in experimental particle physics
how to measure a spin-dependent cross-section.
However, measuring of the spin-dependent part of the forward
scattering amplitude is the complicated challenge.

It was shown in
\cite{2rot}-\cite{8rot} that there is an unambiguous method, which
makes the direct measurement of the real part of the
spin-dependent forward scattering amplitude in the high energy
range possible.
This technique is based on the effect proton (deuteron,
antiproton) beam spin rotation in a polarized nuclear target and
on the deuteron birefringence effect i.e. phenomenon of deuteron
spin rotation and oscillation in a nonpolarized target.
 This technique is based on measurement of the angle of spin rotation of a high energy proton
(deuteron, antiproton) in conditions of transmission experiment.

The similar phenomenon for thermal neutrons was theoretically
predicted in \cite{1rot} and experimentally observed in
\cite{9rot}-\cite{11} (the phenomena of nuclear precession of
neutron spin in a nuclear pseudomagnetic field of a polarized
target).
%
%

When considering particles moving in a storage ring one should
take into account influence of electromagnetic fields, which exist
in the storage ring, on behavior of the particle spin.

According to \cite{14,15}, particle spin behavior in
electromagnetic fields can be described by the
Bargmann-Michel-Telegdi equation.
When the particle has the non-zero electric dipole moment
Bargmann-Michel-Telegdi equation should be supplied with
additional terms, which describe interaction of the EDM with the
electromagnetic field in the storage ring.
This interaction is much weaker than conventional interaction of
the particle spin with the electromagnetic field.
Nevertheless, {the modern experimental technique provides} to
measure such weak interaction of EDM with the electromagnetic
field \cite{6,orlov} and to carry out experimental search of the
deuteron EDM at the level $d \sim 10^{-29} e \cdot cm$.

In the present paper it is shown that modern experimental
technique allows to study particle  spin rotation in a polarized
internal target of a storage ring and, thus, to measure the
spin-dependent part of the forward elastic coherent scattering
amplitude.
It is shown that dynamics of the particle spin in a storage ring
in the experiment with an internal target  can not be completely
described by the BMT equation.
Moreover, it is shown that dynamics of the particle spin in a
storage ring in the experiment with the accuracy necessary for the
EDM search for particles with the spin $S \ge 1$ (deuteron) can
not be completely described by the BMT equation, too.
This is due to the particle possesses the electric and magnetic
tensor polarizabilities.
The equations that can be applied in this case for description of
spin evolution are obtained and additional contributions in these
equations are analyzed.

\section{Particle spin rotation in an electromagnetic field in a storage ring}

Thus, let us consider now a particle with the spin $S$ moving in
the electromagnetic field of a storage ring.
Interaction of the particle magnetic moment with the
electromagnetic field presenting in the storage ring leads to
rotation of the particle spin with respect to the momentum
direction.
To describe particle spin evolution it is used the
Bargmann-Michel-Telegdi equation \cite{14,15} as follows:

%
\begin{equation}
\frac{d\vec{P}}{dt}=[\vec{P}\times\vec{\Omega}],
\label{2.1}
\end{equation}
where $t$ is the time in the laboratory system,
\begin{equation}
\vec{\Omega}=\frac{e}{mc}\left[\left(a+\frac{1}{\gamma}\right)\vec{B}
-a\frac{\gamma}{\gamma+1}\left(\vec{\beta}\cdot\vec{B}\right)\vec{\beta}-
\left(\frac{g}{2}-\frac{\gamma}{\gamma+1}\right)\vec{\beta}\times\vec{E}\right],
\label{2.2}
\end{equation}
$m$ is the mass of the particle, $e$ is its charge, $\vec{P}$ is
the spin polarization vector in the deuteron rest frame, $\gamma$
is the Lorentz-factor,
 $\vec{\beta}=\vec{v}/c$, $\vec{v}$ is the particle velocity, $a=(g-2)/2$, $g$ is the gyromagnetic ratio, $\vec{E}$ and
 $\vec{B}$ are the electric and magnetic fields in the point of
 particle location.

The parameter $a$ for protons is $a=1.79$ and for deuterons
$a=-0.14$. As a result, for the typical field in a storage ring
 ($B \approx 10^2 \div 10^4 ~gauss$) the spin rotation frequency is
 $\Omega \approx 10^5 \div 10^7~sec^{-1}$.

If a particle possesses an intrinsic dipole moment then the
additional term that describes  spin rotation induced by the EDM
should be added to (\ref{2.1}) \cite{6}
\begin{equation}
\frac{d\vec{P}_{edm}}{dt}=\frac{d}{\hbar}
\left[\vec{P}\times\left(\vec{\beta}\times\vec{B}+\vec{E}\right)\right],
\label{2.3}
\end{equation}
where $d$ is the electric dipole moment of a particle.

As a result, evolution of the deuteron spin due to the magnetic
and electric dipole momenta can be described by the following
equation:
\begin{eqnarray}
\frac{d\vec{P}}{dt}=
\frac{e}{mc}\left[\vec{P}\times\left\{\left(a+\frac{1}{\gamma}\right)\vec{B}
-a\frac{\gamma}{\gamma+1}\left(\vec{\beta}\cdot\vec{B}\right)\vec{\beta}-
\left(\frac{g}{2}-\frac{\gamma}{\gamma+1}\right)\vec{\beta}\times\vec{E}\right\}\right]
+d\left[\vec{P}\times\left(c\vec{\beta}\times\vec{B}+\vec{E}\right)\right].
\label{2.4}
\end{eqnarray}
%

The typical frequency $\omega_{d}$ of precession caused by
particle EDM interaction with the electromagnetic field is many
orders less than $\Omega$.
In conditions of the experiments \cite{6,orlov}, which are planned
for the deuteron EDM search at the level $d \approx 10^{-29} e
\cdot cm$, the precession frequency is $\omega_{d} \approx 10^{-8}
\cdot 10^{-9} ~sec^{-1}$.
Nevertheless, according to \cite{6,orlov}, the change of deuteron
spin direction in the storage ring can be measured even when
caused by such small $\omega_d$.

When considering particles with the spin $S \ge 1$, the more so
since the accuracy of the experiments is expected to be high, the
particular attention should be focused on behavior of the particle
spin in the storage ring, because it is strongly influenced by
several additional interactions
\cite{8rot,nastya,birefringence,me}.
Note that a particle with the spin $S \ge 1$ has the electric and
magnetic polarizabilities.
Particle interaction with the electromagnetic field in the storage
ring $\hat{V}_{\vec{E}}$ aroused by its tensor electric
polarizability can be expressed as follows (suppose that $\vec{E}$
and $\vec{B}$ are orthogonal to the particle momentum - this case
is realized in the storage ring):
\begin{equation}
\hat{V}_{\vec{E}}=-\frac{1}{2}\hat{\alpha}_{ik}(E_{eff})_{i}(E_{eff})_{k},
 \label{VE}
\end{equation}
where $\hat{\alpha}_{ik}$ is the electric polarizability tensor of
the particle, $\vec{E}_{eff}=(\vec{E}+\vec{\beta} \times \vec{B})$
is the effective electric field in the point of particle location.
The expression (\ref{VE}) can be rewritten as follows:
\begin{eqnarray}
\hat{V}_{\vec{E}} =
\alpha_{S}E^{2}_{eff}-\alpha_{T}E^{2}_{eff}\left(\vec{S}\vec{n}_{E}\right)^{2},~
\vec{n}_{E}  =
\frac{\vec{E}+\vec{\beta}\times\vec{B}}{|\vec{E}+\vec{\beta}\times\vec{B}|}
 \label{VE1}
\end{eqnarray}
where $\alpha_{S}$ is the scalar electric polarizability and
$\alpha_{T}$ is the tensor electric polarizability of the
particle.

The mentioned interaction results in rotation and oscillations of
the spin of a particle moving in an electric field
\cite{8rot,nastya,birefringence,me}.
The frequency of this rotation $\omega_{TE}$ is determined by the
tensor electric polarizability $\alpha_T$.
Theoretical evaluations \cite{polarizability} give for the
deuteron polarizability $\alpha_T \sim 10^{-40}~cm^3$.
Therefore, for a deuteron in a storage ring the typical energy (in
the frequency units) of spin interaction with the electric field
due to the tensor electric polarizability is about $\omega_{TE}
\sim 10^{-5}$ in the field $E_{eff} \sim 10^4$.

A particle with the spin $S \ge 1$ also has the magnetic
polarizability, which is described by the magnetic polarizability
tensor $\hat{\beta}_{ik}$ and interaction of the particle with the
magnetic field due to the tensor magnetic polarizability is as
follows:
\begin{equation}
\hat{V}_{\vec{B}}=-\frac{1}{2}\hat{\beta}_{ik}(B_{eff})_{i}(B_{eff})_{k},
 \label{VB}
\end{equation}
where $(B_{eff})_{i}$ are the components of the effective magnetic
field $\vec{B}_{eff}=(\vec{B}-\vec{\beta} \times \vec{E})$; the
energy of interaction $\hat{V}_{\vec{B}}$ (\ref{VB}) can be
expressed as:
\begin{equation}
\hat{V}_{\vec{B}}=\beta_{S}B_{eff}^{2}-\beta_{T}B_{eff}^{2}\left(\vec{S}\vec{n}_{B}\right)^{2},~
\vec{n}_B=\frac{\vec{B}-\vec{\beta} \times
\vec{E}}{|\vec{B}-\vec{\beta} \times \vec{E}|}.
 \label{VB1}
\end{equation}
where $\beta_{S}$ is the scalar magnetic polarizability and
$\beta_{T}$ is the tensor magnetic polarizability of the particle.

Similarly interaction $\hat{V}_{\vec{E}}$, interaction
$\hat{V}_{\vec{B}}$ arouses rotation and oscillations of the
particle spin with the frequency $\omega^{\mu}_T$.
According to calculations \cite{polarizability} the tensor
magnetic polarizability can be evaluated as $\beta_T \sim 2 \cdot
10^{-40}~cm^3$.
Therefore, this contribution {provides rotation with the
frequency} $\omega^{\mu}_T \sim 10^{-5}~sec^{-1}$ in the field
$B_{eff} \sim 10^4~gauss$.

As it is obvious the frequencies $\omega_{TE}$ and
$\omega^{\mu}_T$ are much higher than the frequency $\omega_d$,
caused by the deuteron EDM (if it is nonzero). Therefore, the
influence of the tensor polarizabilities on particle spin
evolution in a storage ring should be considered with particular
attention.

\section{The index of refraction and effective potential energy of particles in medium.}

Recall now that a storage ring can contain matter inside (gas, jet
target). Presence of such a target influences on spin behavior
\cite{8rot,nastya,birefringence,me} along with the influence from
the electromagnetic fields.
To make understanding of target influence on spin behavior clear
let us recollect the following.
Close connection between the coherent elastic scattering amplitude
at zero angle $f(0)$ and the refraction index of medium $n$ has
been established as a result of numerous studies (see, for
example, \cite{12, Goldberger}):
\begin{equation}
n=1+\frac{2\pi N }{k^{2}}f\left( 0\right) \label{refr_ind}
\end{equation}
where $N $ is the number of particles per $cm^{3}$, $k$ is the
wave number of the particle incident on the target.

The expression (\ref{refr_ind}) was derived in assumption that
$n-1 \ll 1$. If $k \rightarrow 0$ then $(n-1)$ grows and
expression for $n$ has the form
\[
n^2=1+\frac{4\pi N }{k^{2}}f\left( 0\right) \label{refr_ind2}
\]
Let us consider particle refraction on the vacuum-medium boundary.

The wave number of the particle in vacuum is denoted $k$.
The wave number of the particle in  medium is $k^{\prime} = k n$.
As it is evident the particle momentum in vacuum $p=\hbar k$ is
not equal to the particle momentum in medium.
Therefore, the particle energy in vacuum $E=\sqrt{\hbar^2 k^2
c^2+m^2 c^4}$ is not equal to the particle energy in medium
$E_{med}=\sqrt{\hbar^2 k^2 n^2 c^2+m^2 c^4}$

\begin{figure}[htbp]
\epsfxsize = 5 cm \centerline{\epsfbox{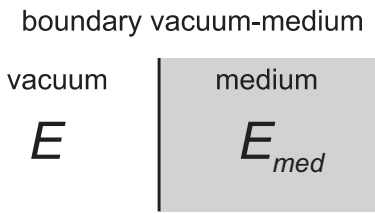}}
\caption{The particle energy in vacuum $E$ is not equal to that in
medium $E_{med}$.}
\end{figure}

The energy conservation law immediately necessitates to suppose
that the particle in medium possesses an effective potential
energy $V_{eff}$ (see the detailed theory in \cite{Goldberger}).
This energy can be easily found  from the evident equality
\[
E={E_{med}}+V_{eff}
\]
i.e.
\begin{equation}
V_{eff}=E-E_{med}=- \frac{2 \pi {\hbar}^2}{m \gamma} {N} f(E,0) =
(2 \pi)^3 N T(E), \label{U1}
\end{equation}
where it is used $f(E,0)=- (2 \pi)^2 ~\frac{E}{c^2 \hbar^2} ~T(E)=
- (2 \pi)^2 ~\frac{m \gamma }{\hbar^2}~ T(E)$, $T(E)$ is the
T-matrix \cite{Goldberger}.

Above we considered the rest target. But in storage rings moving
bunches can be used as a target. Therefore we should generalize
the expressions (\ref{refr_ind},\ref{U1}) for this case. Thus, let
us consider the collision of two bunches of particles. Suppose
that in the rest frame of the storage ring the particles of the
first beam have the energy $E_1$ and Lorentz-factor $\gamma_1$,
whereas particles of the second beam are characterized by the
energy $E_2$ and Lorentz-factor $\gamma_2$. Let us recollect that
the phase of a wave in a medium is Lorentz-invariant. Therefore,
we can find it by the following way. Let us choose the reference
frame, where the second beam rests. As in this frame particles of
the second beam rest, then the refraction index can be expressed
in the conventional form (\ref{refr_ind}):
\begin{equation}
n_1^{\prime }=1+\frac{2\pi N_2^{\prime } }{{k_1^{\prime
}}^{2}}f\left(E_1^{\prime}, 0\right), \label{refr_ind3}
\end{equation}
where $N_2^{\prime }=\gamma_2^{-1} N_2$ is the density of the
bunch 2 in its rest frame and $N_2$ is the density of the second
bunch in the storage ring frame, $k_1^{\prime}$, $E_1^{\prime}$
are the wavenumber and energy of particles of the first bunch in
the rest frame of the bunch 2, respectively. Suppose the length of
the bunch 2 in its rest frame is $L$, then $L=\gamma_2 ~ l$, where
$l$ is the length of this bunch in the storage ring frame.

Now the change of the phase of the wave caused by the interaction
of the particle 1 with the particles of bunch 2 can be found:
\begin{equation}
\phi=k_1^{\prime}(n_1^{\prime}-1)L=\frac{2 \pi
N_2^{\prime}}{k_1^{\prime}} f(E_1^{\prime},0)~L =\frac{2 \pi
N_2}{k_1^{\prime}}{f(E_1^{\prime},0)}{k_1^{\prime}}~l  ,
\label{phase}
\end{equation}

It is known \cite{Goldberger}that the ratio
$\frac{f(E_1^{\prime},0)}{k_1^{\prime}}$ is invariant, therefore,
$\frac{f(E_1^{\prime},0)}{k_1^{\prime}}=\frac{f(E_1,0)}{k_1}$,
where $f(E_1,0)$ is the amplitude of elastic coherent forward
scattering of the particle 1 by the moving particle 2 in the rest
frame of the storage ring.

As a result
\begin{equation}
\phi=\frac{2 \pi N_2}{k_1} f(E_1,0)~l=\frac{2 \pi N_2}{k_1}
f(E_1,0)~v_{rel}~t, \label{phase1}
\end{equation}
where $v_{rel}$ is the velocity of relative motion of the particle
1 and bunch 2 (for opposing motion $v_{rel}=(v_1+v_2)(1+\frac{v_1
v_2}{c^2})^{-1}$), $t$ is the time of interaction of the particle
1 with the bunch 2 in the rest frame of the storage ring.

The particle with the velocity $v_1=\frac{\hbar k_1 c^2}{E_1}$
passes the distance $z=v_1~t$ over the time $t$. It should be
noted that the path length $z$ differs from the length of the
bunch 2, because it moves. Expression (\ref{phase1}) can be
rewritten as:
\begin{equation}
\phi=\frac{2 \pi N_2}{k_1} f(E_1,0)~\frac{v_{rel}}{v_1}~z=k_1
(n_1-1)z, \label{phase_z}
\end{equation}
where the index of refraction of the particle 1 by the beam of
moving particles 2 is:
\begin{equation}
n_1=1+\frac{2 \pi {N}_2}{{k_1}^2}~\frac{v_{rel}}{v_1}f(E_1,0)
\label{r_ind_z}
\end{equation}
When $v_2=0$, the conventional expression {(\ref{refr_ind})}
follows from (\ref{r_ind_z}).

Now the effective potential energy $V_{eff}$ being acquired by a
particle 1 when colliding with the particles of bunch 2 can be
found:
\begin{equation}
V_{eff}=E_1-E_{1~med}=E_1-\sqrt{p_1^2 c^2 n_1^2 + m_1^2 c^4} = -2
\pi \hbar^2 N_2  v_{rel} \frac{f(E_1,0)}{p_1}=-2 \pi \hbar^2 N_2
v_{rel} \frac{f(E_1^{\prime},0)}{p_1^{\prime}}, \label{Veff_new}
\end{equation}
therefore
\begin{equation}
V_{eff}=- \frac{2 \pi \hbar^2 N_2 }{m_1 \gamma_1 \gamma_2}
f(E_1^{\prime},0)= (2 \pi)^3 N_2 T (E_1^{\prime}),
\label{Veff_new1}
\end{equation}
where $E_1^{\prime}=m_1 c^2 \gamma_1 \gamma_2$ is the energy of
the particle 1 in the rest frame of the bunch 2, $p_1$ is the
particle 1 momentum in the storage ring frame, while
$p_1^{\prime}$ is the particle 1 momentum in the rest frame o the
bunch 2, $p_1^{\prime}=\frac{E_1^{\prime} v_1^{\prime}}{c^2}$, it
is taken into account that $v_1^{\prime}=v_{rel}$.

When obtaining (\ref{Veff_new}) it was used $|n_1-1| \ll 1$.


Let us consider now what happens when the {particle possesses a
spin}.
In this case the amplitude of the zero-angle scattering depends on
the particle spin and, as a consequence, the index of refraction
depends on the particle spin and can be written as:

\begin{equation}
\hat {n}_1=1+\frac{2 \pi {N}_2}{{k_1}^2}~\frac{v_{rel}}{v_1}\hat
{f}(E_1,0) \label{hatn1},
\end{equation}
where $\hat {f}\left(E_1,{0} \right) = Tr \hat {\rho} _{J} \hat
{F}\left( {0} \right)$,  $\hat {\rho} _{J} $ is the spin density
matrix of the scatterers, $\hat {F}\left( {0} \right)$ is the
operator of the forward scattering amplitude, acting in the
combined spin space of the particle and scatterer spin $\vec {J}$.

According to the above (see (\ref{U1},\ref{Veff_new1})) a particle
possesses some effective potential energy $V$ in matter.
If the amplitude $ \hat {f}(0)$ of particle scattering depends on
the particle spin, then the effective energy depends on the spin
orientation \cite{2rot}-\cite{8rot}:
\begin{equation}
\hat{V}_{eff} =- \frac{2 \pi \hbar^2 N_2 }{m_1 \gamma_1 \gamma_2}
\hat{f}(E_1^{\prime},0).  \label{1.1}
\end{equation}

Thus, for example, the amplitude $ \hat {f}(0)$ of the spin $S=1$
particle (for example, deuteron) scattering even in an unpolarized
target depends on the particle spin and can be written as:
\begin{equation}
\label{hatf} \hat{f}(E_1^{\prime},0) = d + d_{1} (\vec{S}
\vec{n})^{2}.
\end{equation}
where $\vec{S}$ is the deuteron spin operator, $\vec{n}$ is the
unit vector along the deuteron momentum $\vec{k}$.

Substituting the expression for $\hat{f(0)}$ (\ref{hatf}) in
(\ref{1.1}) one can obtain for a particle with the spin
 $S=1$:
\begin{equation}
\hat{V}_{eff}=-\frac{2\pi \hbar^{2}}{m_1 \gamma_1 \gamma_2} N_2
\left(d+d_{1}\left(\vec{S}\vec{n}\right)^{2}\right). \label{1.2}
\end{equation}

Let the quantization axis z is directed along $\vec{n}$ and $M$
denotes the magnetic quantum number. Then, for a particle in a
state that is the eigenstate of the operator $S_{z}$ of spin
projection onto the z-axis, the effective potential energy can be
written as:
\begin{equation}
\hat{V}_{eff}=-\frac{2\pi \hbar^{2}}{m_1 \gamma_1 \gamma_2}N_2
\left(d+d_{1}M^{2}\right). \label{1.3}
\end{equation}

 According to (\ref{1.3}) splitting of deuteron energy levels
in matter is similar to splitting of atom energy levels in the
electric field aroused from the quadratic Stark effect.
 Therefore,
the above effect could be considered as caused by splitting of the
spin levels of the particle in the pseudoelectric nuclear field of
matter.

{Comparison of (\ref{1.2}) with (\ref{VE1}) provides to conclude
that interactions (\ref{1.2}) and (\ref{VE1}) are similar,
therefore
 we can observe the effect of spin rotation and oscillations
for a particle with $S \ge 1$ passing through nonpolarized matter
(birefringence effect) \cite{5rot,6rot}.}
Henceforth, for the expression (\ref{1.2}) we will use the
notation $\hat{V}_{E}^{nucl}=\hat{V}_{eff}$.


\section{Spin rotation of proton (deuteron, antiproton) in a storage ring with a polarized target}

Let us consider now the experiments deals with the use of
polarized beams and targets.
In this case the amplitude of the elastic coherent scattering at
the zero angle depends on  the vector polarization $\vec{P}_t$ of
the target nuclei (if the spin of the target nuclei $J \ge 1$,
then the addition depending on the target tensor polarization also
appears, but this addition will be omitted in below
considerations, the general case is considered in
\cite{2rot,6rot}).

For the sake of concreteness suppose the target to be rest
($\gamma_2=1$, $\gamma_1=\gamma$).

The contribution to the amplitude $\hat{f(0)}$ proportional to the
vector polarization of the target nuclei can be expressed as:

\begin{equation}
\hat{f}(\vec{P}_t,0)=A_1 \vec{S} \vec{P}_t + A_2
(\vec{S}\vec{n})(\vec{n} \vec{P}_t). \label{fPt}
\end{equation}
Contributions from weak P,T-odd interactions (see \cite{2rot},
\cite{nastya}-\cite{me}) are omitted here.

Correspondingly, the contribution to the effective potential
energy of particle interaction with matter caused by polarization
of target nucleus spins looks like:
\begin{equation}
\hat{V}_{eff}(\vec{P}_t)=-\frac{2 \pi \hbar^2}{m \gamma} N (A_1
\vec{S} \vec{P}_t + A_2 (\vec{S}\vec{n})(\vec{n} \vec{P}_t)).
\label{Veff}
\end{equation}

The expression $\hat{V}_{eff}(\vec{P}_t)$ can be rewritten as:
\begin{equation}\label{Veff2}
\hat{V}_{G}^{nucl} \equiv \hat{V}_{eff}(\vec{P}_t)=-\vec{\mu}
\vec{G}=-\frac{\mu}{S} \vec{S} \vec{G}
\end{equation}
where $\mu$ is the particle magnetic moment,
\begin{equation}
\vec{G}=\frac{2 \pi \hbar^2 S}{m \gamma \mu} N (A_1 \vec{P}_t +
A_2 \vec{n}(\vec{n} \vec{P}_t)). \label{26}
\end{equation}

Recall now that the energy of interaction of the magnetic moment
$\vec{\mu}$ with a magnetic field $\vec{B}$ is as follows:
\begin{equation}
V_{mag}=-\vec{\mu} \vec{B}=-\frac{\mu}{S} \vec{S} \vec{B}.
\label{Vmag}
\end{equation}

Comparing (\ref{Veff2}) and (\ref{Vmag}) one can easily find they
are identical.
Therefore, $\vec{G}$ can be interpreted as the effective
pseudomagnetic field acting on the spin of the particle in matter
with polarized nuclei.
Pseudomagnetic field $\vec{G}$ is caused by nuclear interaction of
particles with scatterers.
Similar particle spin precession in conventional magnetic field,
spin also precesses in the field $\vec{G}$  (this phenomenon is
called nuclear precession of the particle spin).
This phenomenon was described for the first time for slow neutrons
in \cite{1rot} and observed in \cite{9rot}-\cite{11}.

\section{The equations describing spin evolution of a particle in a storage ring}

\subsection{Interactions contributing to the spin motion of a
particle in a storage ring}

Thus, according to the above analysis,  when considering evolution
of the spin of a particle in a storage ring one should take into
account several interactions:

1. interactions of the magnetic and electric dipole moments with
electromagnetic fields;

2. interaction of the particle with electric fields due to the
tensor electric polarizability;

3. interaction of the particle with magnetic fields due to the
tensor magnetic polarizability;

4. interaction of the particle with the pseudoelectric nuclear
field of matter;

5. interaction of the particle with the pseudomagnetic nuclear
field of polarized matter.

The equation for the particle spin wavefunction considering all
these interactions is as follows:
\begin{equation}
i\hbar\frac{\partial\Psi(t)}{\partial
t}=\left(\hat{H}_{0}+\hat{V}_{EDM}+\hat{V}_{\vec{E}}+\hat{V}_{\vec{B}}+\hat{V}_E^{nucl}+
\hat{V}_G^{nucl}\right)\Psi(t) \label{1}
\end{equation}
where $\Psi(t)$ is the particle spin wavefunction,

{$\hat{H}_{0}$ is the Hamiltonian describing spin behavior caused
by interaction of the magnetic moment with the electromagnetic
field (equation (\ref{1}) with the only $\hat{H}_{0}$  summand
converts to the Bargmann-Michel-Telegdi equation);}

$\hat{V}_{EDM}$ describes interaction of the particle EDM $d$ with
the electric field;

$\hat{V}_{\vec{E}}$ describes interaction of the particle with the
electric field due to the tensor electric polarizability;

$\hat{V}_{\vec{B}}$ is the energy of interaction of the particle
with the magnetic field due to the tensor magnetic polarizability;

$\hat{V}_E^{nucl}$ describes the effective potential energy of
particle interaction with the pseudoelectric field of the target
(\ref{1.2}).

$\hat{V}_G^{nucl}$ describes the effective potential energy of
particle interaction with the pseudomagnetic field of the target
(\ref{Veff2}).

\subsection{Deuteron spin rotation in electromagnetic fields of a storage ring}

Let us consider a deuteron moving in a storage ring with low
pressure of the residual gas ($10^{-10}$ Torr) and without targets
inside the storage ring.
In this case we can omit the effects caused by the interactions
$\hat{V}_E^{nucl}$ and $\hat{V}_G^{nucl}$.
 According to the above analysis spin
behavior of a deuteron, nevertheless, can not be described by the
Bargmann-Michel-Telegdi equation.
According to the above interactions $\hat{V}_E$ and $\hat{V}_B$
should be considered when describing spin evolution.
 The corresponding equations
can be written as follows \cite{nastya}-\cite{me}:
\begin{eqnarray}
\left\{
\begin{array}{l}
\frac{d\vec{P}}{dt}= \left[\vec{P}\times
\Omega(\vec{d}) \right]-\\
-\frac{2}{3}\frac{\alpha_{T}E^{2}_{eff}}{\hbar}[\vec{n}_{E}\times\vec{n}_{E}^{\prime}]
-\frac{2}{3}\frac{\beta_{T}B^{2}_{eff}}{\hbar}[\vec{n}_{B}\times\vec{n}_{B}^{\prime}],\\
{} \\
\frac{dP_{ik}}{dt}  =
-\left(\varepsilon_{jkr}P_{ij}\Omega_{r}(d)+\varepsilon_{jir}P_{kj}\Omega_{r}(d)\right)
 - \\
-
\frac{3}{2}\frac{\alpha_{T}E^{2}_{eff}}{\hbar}\left([\vec{n}_{E}\times\vec{P}]_{i}n_{E,\,k}
+n_{E,\,i}[\vec{n}_{E}\times\vec{P}]_{k}\right)-\\
 -
\frac{3}{2}\frac{\beta_{T}B^{2}_{eff}}{\hbar}\left([\vec{n}_{B}\times\vec{P}]_{i}n_{B,\,k}
+n_{B,\,i}[\vec{n}_{B}\times\vec{P}]_{k}\right),
\\
\end{array}
\right. \label{50}
\end{eqnarray}
where
\begin{eqnarray}
\begin{array}{l}
\vec{\Omega}(d)  =  \vec{\Omega} + \vec{\Omega}_{d}, \nonumber \\
\vec{\Omega}  =
\frac{e}{mc}\left\{\left(a+\frac{1}{\gamma}\right)\vec{B} -
\left(\frac{g}{2}-\frac{\gamma}{\gamma+1}\right)\vec{\beta}\times\vec{E}\right\}, \nonumber \\
\vec{\Omega}_{d}  =
\frac{d}{\hbar}\left({\vec{E}}+\vec{\beta}\times\vec{B}\right),\\
P_{ik}=\textrm{Sp} ~\rho ~\hat{Q}_{ik},\\
\hat{Q}_{ik}=\frac{3}{2} \left[ S_i S_k + S_k S_i - \frac{4}{3}
\delta_{ik}\right]
\label{2.24}
\end{array}
\end{eqnarray}
$\rho$ is the spin density matrix of the particle, $m$ is the
particle mass, $e$ is its charge, $\vec{P}$ is the spin
polarization vector, $P_{ik}$ is the spin polarization tensor,
$P_{xx}+P_{yy}+P_{zz}=0$, $\gamma$ is the Lorentz-factor,
 $\vec{\beta}=\vec{v}/c$,
$\vec{v}$ is the particle velocity, $a=(g-2)/2$, $g$ is the
gyromagnetic ratio, $ \vec{E}$ and
 $\vec{B}$ are the electric and magnetic fields in the point of
 particle location, $\vec{E}_{eff}=(\vec{E}+\vec{\beta} \times
\vec{B})$, $\vec{B}_{eff}=(\vec{B}-\vec{\beta} \times \vec{E})$,
$\vec{n}_{E}=\frac{\vec{E}+\vec{\beta}\times\vec{B}}{|\vec{E}+\vec{\beta}\times\vec{B}|}$,
$\vec{n}_B=\frac{\vec{B}-\vec{\beta} \times
\vec{E}}{|\vec{B}-\vec{\beta} \times \vec{E}|}$,
 $n_{E,\,i}^{\prime}=P_{ik}n_{E,\,k}$,
$n_{Bi}^{\prime}=P_{il}n_{Bl}$, $\Omega_{r}(d)$ are the components
of the vector $\vec{\Omega}(d)$ ($r=1,2,3$ correspond to $x,y,z$,
respectively).
The equations for particle spin motion (\ref{50}) can be rewritten
as follows:
\begin{eqnarray}
\frac{d\vec{P}}{dt}=[\vec{P}\times\vec{\Omega}(d)]+\Omega_T[\vec{n}_E\times\vec{n}_E^{\prime}]+
\Omega_T^{\mu}[\vec{n}_B\times\vec{n}_B^{\prime}], \nonumber\\
{} \nonumber\\
\frac{d\vec{P_{ik}}}{dt}=-(\epsilon_{jkr}P_{ij}\Omega_r(d)+\epsilon_{jir}P_{kj}\Omega_r(d))+
\Omega_T^{\prime}([\vec{n}_E\times\vec{P}]_i
n_{Ek}+n_{Ei}[\vec{n}_E\times\vec{P}]_k) + \nonumber \\
{} \nonumber\\
+\Omega_T^{\prime \mu}([\vec{n}_B\times\vec{P}]_i
n_{Bk}+n_{Bi}[\vec{n}_B\times\vec{P}]_k)\label{BMT} \label{BMT+}
\end{eqnarray}
 where
\begin{eqnarray}
\begin{array}{l}
\Omega_T=-\frac{2}{3} \frac{\alpha_T E_{eff}^2}{\hbar} ,
~~\Omega_T^{\prime}=-\frac{3}{2} \frac{\alpha_T E_{eff}^2}{\hbar},
~~\Omega_T^{\prime}=-\frac{2}{3} \Omega_T,\nonumber \\
{}\nonumber \\
 \Omega_T^{\mu}=-\frac{2}{3} \frac{\beta_T
B_{eff}^2}{\hbar} , ~~\Omega_T^{\prime \mu}=-\frac{3}{2}
\frac{\beta_T B_{eff}^2}{\hbar}, ~~\Omega_T^{\prime
\mu}=-\frac{2}{3} \Omega_T^{\mu}.
\end{array}
\end{eqnarray}

Suppose the external electric field in the storage ring
$\vec{E}=0$ and the particle moves  {along the circle orbit}.

Let us now consider the equation (\ref{BMT+}) in the coordinate
system that rotates with the frequency of particle velocity
rotation.
 {In such a system spin rotates with respect to the momentum
with the frequency determined by $(g-2)$.}%
The coordinate system and vectors $\vec{v},\vec{E}, \vec{B}$ are
shown in figure and denote the axes by $x,y,z$ (or $1,2,3$,
respectively).
\begin{figure}[!h]
\begin{center}
\includegraphics[width=8cm,keepaspectratio]{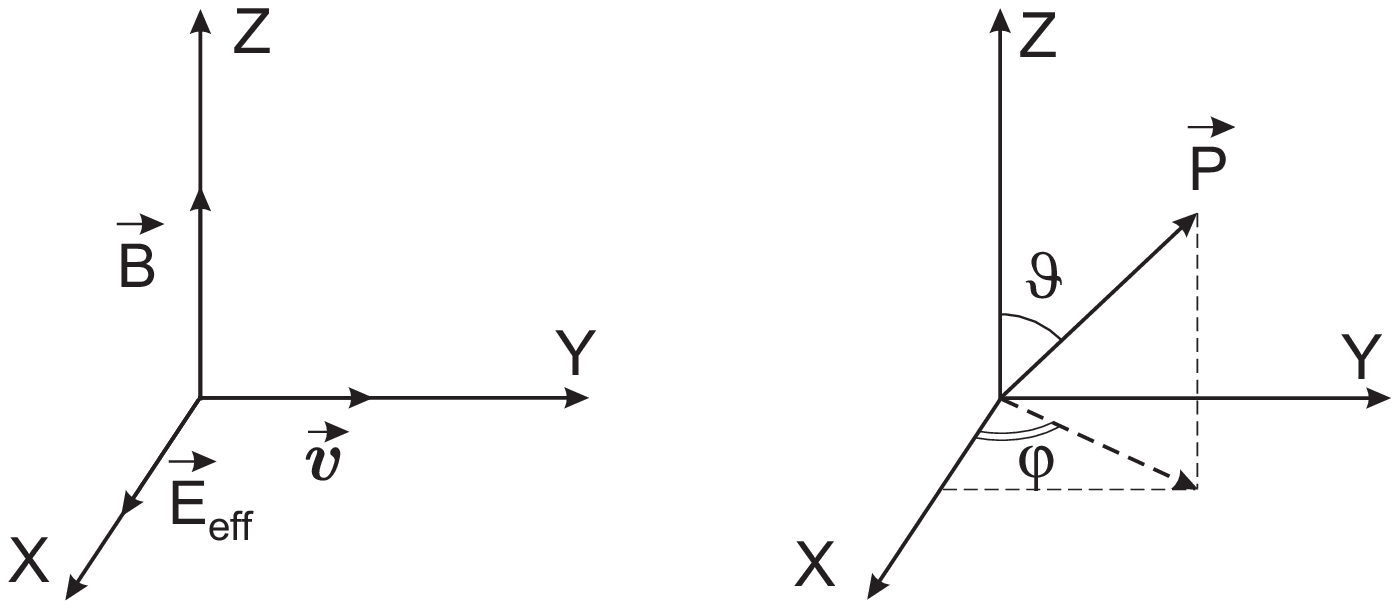}
\caption{} \label{coordinate}
\end{center}
\end{figure}

As a result, the system (\ref{50}) can be written as:
\begin{eqnarray}
\begin{array}{l}
\frac{dP_1}{dt}=\Omega P_2-\Omega_T^{\mu} P_{23},\\
{}  \\
 \frac{dP_2}{dt}=-\Omega P_1+(\Omega_T^{\mu}-\Omega_T)P_{13}+\omega_d P_3,\\
 {}  \\
 \frac{dP_3}{dt}=\Omega_T P_{12}-\omega_d P_2,
 \end{array}
\label{Pi}
\end{eqnarray}
where $\omega_d=\frac{dE_1^{eff}}{\hbar}$,
\begin{eqnarray}
\begin{array}{l}
\frac{dP_{11}}{dt} = 2 \Omega P_{12}+2 \omega_d P_{23},  \\
 {}  \\
\frac{dP_{22}}{dt} = -2 \Omega P_{12}, \\
 {}  \\
\frac{dP_{33}}{dt} =-2 \omega_d P_{23}, \\
 \end{array}
\label{sys1}
\end{eqnarray}
\begin{eqnarray}
\begin{array}{l}
\frac{dP_{12}}{dt} = -\Omega \left(
P_{11}-P_{22}\right)-\Omega_T^{\prime} P_3+\omega_d P_{13},
 \\
 {}  \\
\frac{dP_{13}}{dt} = \Omega P_{23}+{\Omega}_T^{\prime}P_2-{\Omega}_T^{\prime \mu}P_2 -\omega_d P_{12},  \\
{}  \\
\frac{dP_{23}}{dt} = -\Omega P_{13}+{\Omega}_T^{\prime \mu}P_1
-\omega_d (P_{22}-P_{33}).
 \end{array}
\label{p12def}
\end{eqnarray}
Remember that $P_{11}+P_{22}+P_{33}=0$ and $P_{ik}=P_{ki}$.

\subsection{Contribution from the EDM and tensor polarizabilities
to deuteron spin oscillation}

Let us consider the system (\ref{Pi}-\ref{p12def}) more
attentively.

Suppose that deuteron has neither EDM no tensor polarizabilities:
in this case the system (\ref{Pi}-\ref{p12def}) can be expressed:
\begin{eqnarray}
\begin{array}{l}
\frac{dP_1}{dt}=\Omega P_2,~
 \frac{dP_2}{dt}=-\Omega P_1,~
 \frac{dP_3}{dt}=0,
 \end{array}
\label{Pi0}
\end{eqnarray}
\begin{eqnarray}
\begin{array}{l}
\frac{dP_{11}}{dt} = 2 \Omega P_{12},  ~ \frac{dP_{22}}{dt} = -2
\Omega P_{12}, ~ \frac{dP_{33}}{dt} =0,
 \end{array}
\label{sys10}
\end{eqnarray}
\begin{eqnarray}
\begin{array}{l}
\frac{dP_{12}}{dt} = -\Omega \left( P_{11}-P_{22}\right),
 ~
\frac{dP_{13}}{dt} = \Omega P_{23},  ~ \frac{dP_{23}}{dt} =
-\Omega P_{13}.
 \end{array}
\label{p12def0}
\end{eqnarray}
This is the conventional system of BMT equations that describes
particle spin rotation with the frequency equal to $(g-2)$
precession frequency ${\Omega}=\frac{ea}{mc}{B}$.
The component $P_3$ of vector polarization in this conditions is
equal to a constant $(\frac{dP_{3}}{dt} =0)$ along with the
component $P_{33}$ of tensor polarization  $(\frac{dP_{33}}{dt}
=0)$.

Suppose the deuteron EDM differs from zero. Then the above system
of equations converts to:
\begin{eqnarray}
\begin{array}{l}
\frac{dP_1}{dt}=\Omega P_2,~
 \frac{dP_2}{dt}=-\Omega P_1+\omega_d P_3,~
 \frac{dP_3}{dt}=-\omega_d P_2,
 \end{array}
\label{PiE}
\end{eqnarray}
\begin{eqnarray}
\begin{array}{l}
\frac{dP_{11}}{dt} = 2 \Omega P_{12}+2 \omega_d P_{23},  ~
\frac{dP_{22}}{dt} = -2 \Omega P_{12}, ~ \frac{dP_{33}}{dt} =-2
\omega_d P_{23},
 \end{array}
\label{sys1E}
\end{eqnarray}
\begin{eqnarray}
\begin{array}{l}
\frac{dP_{12}}{dt} = -\Omega \left( P_{11}-P_{22}\right)+\omega_d
P_{13},
 ~
\frac{dP_{13}}{dt} = \Omega P_{23} -\omega_d P_{12},  ~
\frac{dP_{23}}{dt} = -\Omega P_{13} -\omega_d (P_{22}-P_{33}).
 \end{array}
\label{p12defE}
\end{eqnarray}
From  (\ref{PiE}-\ref{p12defE}) it follows that presence of the
nonzero EDM makes the vertical component $P_3$ of vector
polarization oscillating with the frequency of $(g-2)$ precession
$\Omega$.

According to the idea \cite{orlov} these oscillations can be
eliminated by modulation of the deuteron velocity with the
frequency $\Omega$:
\begin{equation}
v=v_0+\delta v \sin{(\Omega t + \varphi_f)} \label{v}
\end{equation}
here $\varphi_f$ is the phase of forced velocity oscillations.

As  $E_{eff}$ depends on $\vec{\beta}=\vec{v}/c$, it also appears
modulated:
\begin{equation}
E_{eff}=E_{eff}^0+\delta E_{eff} \sin{(\Omega t + \varphi_f)}
\label{Eeff}
\end{equation}
Therefore $\omega_d=\frac{d E_{eff}}{\hbar}$ is also modulated
with the same frequency. This makes the product $\omega_d P_2 \sim
\sin^2 (\Omega t+ \varphi_f)$.
Therefore, averaging this value over the period of $(g-2)$
precession gives the result time-independent (i.e.
$\frac{dP_3}{dt}=const$) and $P_3 (t)=P_3 (0)+const \cdot t$.
For better measurement conditions it is important to make $P_3
(0)=0$.
 {This is the reason to chose particle spin laying in the
horizontal plane.}

All the above reasoning makes $\frac{dP_{33}}{dt} \sim const$,
too.
Therefore, $P_{33}$ also linearly grows with time $P_{33}
(t)=P_{33} (0)+const \cdot t$.
However, when the spin lays in the horizontal plane $P_{33}(0) \ne
0$.

It is important to note (see below the section 5.5) that choosing
spin orientation corresponding to $\cos^2 \vartheta = \frac{1}{3}$
($\cos \vartheta =\sqrt{\frac{1}{3}}, ~ \sin \vartheta =
\sqrt{\frac{2}{3}}$) one gets the component $P_{33} (0) = 0$,
while $P_3(0) \ne 0$.

Therefore taking $\vartheta$ making $\cos \vartheta=
\sqrt{\frac{1}{3}}$ we can use the component $P_{33}$ for EDM
measurements, too.


Let us consider now the contribution from the electric and
magnetic tensor polarizabilities. Then instead of the system
(\ref{PiE}-\ref{p12defE}) we should consider the system
(\ref{Pi}-\ref{p12def})

Some interesting implications follow from (\ref{Pi}-\ref{p12def}).
 As it was already mentioned above in the experiments for EDM
search it is planned to measure growth of the vertical component
of the polarization vector $P_3$.

According to (\ref{Pi}) time dependence of the vertical component
of the vector polarization $P_3$ is described by the equation
\begin{equation}
\frac{dP_3}{dt}=\Omega_T P_{12}-\omega_d P_2
\end{equation}
As it can be seen from (\ref{Pi}) dependence of $P_3$ on time is
determined by both the EDM and tensor polarizability of deuteron.
It is interesting that the derivative of the tensor polarization
component $P_{33}$ does not contain contributions from the tensor
electric polarizability and is proportional to the EDM only:
\begin{equation}
\frac{dP_{33}}{dt} =-2 \omega_d P_{23}
\end{equation}
Therefore, it is  important to measure the component $P_{33}$,
too.
According to the above spin orientation for this case is
determined by the condition  $\cos^2 \vartheta = \frac{1}{3}$.

\subsection{Contribution from the tensor magnetic polarizability to deuteron spin oscillation}

Contributions to spin rotation and oscillations from the EDM and
polarizabilities are small. Therefore, they, being analyzed, could
be considered as perturbations to the full system
(\ref{Pi}-\ref{p12def}) and the role of each could be studied
separately.

The system of equations taking into account the contribution from
the tensor magnetic polarizability $\beta_T$ is as follows:
\begin{eqnarray}
\begin{array}{l}
\frac{dP_1}{dt}=\Omega P_2-\Omega_T^{\mu} P_{23},~
 \frac{dP_2}{dt}=-\Omega P_1+\Omega_T^{\mu} P_{13},~
 \frac{dP_{13}}{dt} = \Omega P_{23}-{\Omega}_T^{\prime \mu}P_2,  ~
\frac{dP_{23}}{dt} = -\Omega P_{13}+{\Omega}_T^{\prime \mu}P_1
\end{array} \label{sys2}
\end{eqnarray}

Introducing new variables $P_{+}=P_1+iP_2$ and
$G_{+}=P_{13}+iP_{23}$ and recomposing equations (\ref{sys2}) to
determine $P_{+}$ and $G_{+}$ one obtains:
\begin{eqnarray}
\begin{array}{l}
\frac{dP_+}{dt}=-i\Omega P_+ + i\Omega_T^{\mu} G_+,\\
{} \nonumber \\
 \frac{dG_+}{dt}=-i\Omega G_+ + i\Omega_T^{\prime \mu} P_+,\\
\end{array}
\label{sys22}
\end{eqnarray}

or

\begin{eqnarray}
\begin{array}{l}
i\frac{dP_+}{dt}=\Omega P_+ - \Omega_T^{\mu} G_+,\\
{} \nonumber \\
i \frac{dG_+}{dt}=\Omega G_+ - \Omega_T^{\prime \mu} P_+,\\
\end{array}
\label{sys3}
\end{eqnarray}

Let us search $P_+,G_+ \sim e^{i\omega t}$, then (\ref{sys3})
transforms as follows:

\begin{eqnarray}
\begin{array}{l}
\omega \tilde{P}_+=\Omega \tilde{P}_+ - \Omega_T^{\mu} \tilde{G}_+,\\
{} \nonumber \\
\omega \tilde{G}_+=\Omega \tilde{G}_+ - \Omega_T^{\prime \mu}
\tilde{P}_+.
\end{array}
\label{sys33}
\end{eqnarray}
The solution of this system can be easily found:
\begin{eqnarray}
\begin{array}{l}
(\omega -\Omega)^2- \Omega_T^{\mu} \Omega_T^{\prime \mu}=0\\
\end{array}
\label{sys34}
\end{eqnarray}
 that finally gives
 \begin{equation}
\omega_{1,2}=\Omega \pm \sqrt{\Omega_T^{\mu} \Omega_T^{\prime
\mu}}.
\end{equation}
The solution can be rewritten as:
 \begin{equation}
P_+(t)=c_1 e^{-i\omega_1 t}+c_2 e^{-i\omega_2 t}= |c_1|
e^{-i(\omega_1 t-\delta_1)}+|c_2| e^{-i(\omega_2 t-\delta_2)}
\label{P+}
\end{equation}
Therefore,
 \begin{equation}
P_1(t)=|c_1| \cos (\omega_1 t - \delta_1)+|c_2| \cos (\omega_2 t -
\delta_2) \label{reP+}
\end{equation}
 \begin{equation}
P_2(t)=-|c_1| \sin (\omega_1 t - \delta_1)-|c_2| \sin (\omega_2 t
- \delta_2) \label{imP+}
\end{equation}


According to (\ref{reP+},\ref{imP+}) the nonzero deuteron tensor
magnetic polarizability makes the spin rotating with two
frequencies $\omega_1$ and
 $\omega_2$ instead of $\Omega$ and, therefore, experiences beating with the frequency
$\Delta \omega=\omega_1-\omega_2=2 \sqrt{\Omega_T^{\mu}
\Omega_T^{\prime \mu}}=\frac{\beta_T B_{eff}^2}{\hbar} $.

Let us recall now that EDM interaction with the electric field
makes the deuteron spin rotating around the direction of this
field and leads to appearance of $P_3$ component proportional to
$P_2(t)$:
 \begin{equation}
 \frac{dP_3}{dt} \sim - \omega_d P_2.
\end{equation}

Therefore,

 \begin{equation}
 \frac{dP_3}{dt}=\omega_d \left( |c_1| \sin (\omega_1 t - \delta_1)+|c_2| \sin (\omega_2
t - \delta_2) \right).
\end{equation}

According to the idea \cite{orlov} to measure the EDM one should
modulate the particle velocity ($v=v_0+\delta v \sin{(\Omega_f t +
\varphi_f)}$) with the frequency $\Omega_f$ close to the frequency
$\Omega$ of $(g-2)$ precession.

If the magnetic polarizability is equal to zero, then $\omega_1 =
\omega_2=\Omega$ and spin rotates in the horizontal plane with the
frequency $\Omega$. In this case velocity modulation with the same
frequency $\Omega_f=\Omega$ gives:
 \begin{equation}
 \frac{dP_3}{dt} \sim \sin^2 (\Omega t)
\end{equation}
and the vertical component $P_3=\int_0^t \frac{dP_3}{dt'}dt'$
linearly grows with time.

However, $\omega_1 \ne \omega_2$ and velocity modulation, for
example, with the frequency $\Omega=\omega_1$ provides for slow
spin oscillation with the frequency $\omega_1 - \omega_2$ instead
of linear growth.

According to evaluation \cite{polarizability} the tensor magnetic
polarizability $\beta_T \sim 2 \cdot 10^{-40}$, therefore in the
field $B \sim 10^4$ gauss the beating frequency $\Delta \omega
\sim 10^{-5}$.

Measurement of the frequency of this beating makes possible to
measure the tensor magnetic polarizability of the deuteron
(nuclei).

Thus, due to presence of the tensor magnetic polarizability the
horizontal component of spin rotates around $\vec{B}$ with two
frequencies $\omega_1,~\omega_2$ instead of expected rotation with
the frequency $\Omega$.

This is the reason for the component $P_3$ to experience the
similar oscillations caused by the EDM.
Therefore, particle velocity modulation with the frequency
$\Omega$  provides for eliminating of oscillation with the
frequency $\Omega$, but $P_3$ oscillations with the frequency
$\Delta \omega$ rest (similarly $P_{33}$).
Study of these oscillations is necessary because they can distort
the EDM measurements.

\subsection{Contribution from the tensor electric polarizability to deuteron spin oscillation}

Let us consider now contribution caused by the tensor electric
polarizability.
From the system (\ref{sys1}) it follows

\begin{eqnarray}
\begin{array}{l}
\frac{d(P_{11}-P_{22})}{dt} = 4\Omega P_{12},
\\
 {}  \\
\frac{d^2P_{12}}{dt^2} = -\Omega
\frac{d(P_{11}-P_{22})}{dt}-{\Omega}_T^{\prime} \frac{dP_3}{dt}=
-(4\Omega^2+\Omega_{T}{\Omega}_T^{\prime})P_{12}.  \\
 \end{array}
\label{p11-p22}
\end{eqnarray}

Thus we have the equation
\begin{equation}
\frac{d^2P_{12}}{dt^2}+ \omega_{12}^2 P_{12}=0
\end{equation}
where $\omega_{12}=\sqrt{4\Omega^2+\Omega_{T}{\Omega}_T^{\prime}}
\approx 2 \Omega$, because $\Omega_{T}{\Omega}_T^{\prime} \ll
\Omega^2$.

The solution for this equation can be found in the form:
\begin{equation}
P_{12}=c_1 \cos{\omega_{12} t}+c_2 \sin{\omega_{12} t} \label{P12}
\end{equation}

Let us find coefficients $c_1$ and $c_2$: when $t=0$ the equation
(\ref{P12}) gives $c_1=P_{12}(0)$. The coefficient $c_2$ can be
found from
\begin{equation}
\frac{d(P_{12})}{dt}(t \rightarrow 0)=\omega_{12} c_2,
\end{equation}
therefore
\begin{equation}
c_2= \frac{1}{\omega_{12}} \frac{d(P_{12})}{dt}(t \rightarrow 0) ,
\end{equation}
From the equation (\ref{p12def})
\begin{equation}
\frac{dP_{12}}{dt} (t \rightarrow 0)= -\Omega \left( P_{11}(t
\rightarrow 0)-P_{22}(t \rightarrow 0)\right), \label{c2}
\end{equation}
it follows that
\begin{equation}
c_2=-\frac{P_{11}-P_{22}}{2}, \label{c2a}
\end{equation}
and
\begin{equation}
P_{12}=P_{12}(0) \cos{\omega_{12} t}-\frac{P_{11}-P_{22}}{2}
\sin{\omega_{12} t} \label{P12final}
\end{equation}

As a result, one can write the following equation for the vertical
component of the spin $P_3$:
\begin{equation}
\frac{dP_{3}}{dt}=\Omega_T P_{12}(t)= \Omega_T [ P_{12}(0)\cos{2
\Omega t}-\frac{P_{11}(0)-P_{22}(0)}{2} \sin{2\Omega t} ]
\label{P12final1}
\end{equation}

As it can be seen the vertical component of the spin oscillates
with the frequency $2 \Omega$.

The equation describing contribution  from the tensor electric
polarizability and EDM to $P_3$ looks like (\ref{Pi}):
\begin{equation}
\frac{dP_3}{dt}=\Omega_T P_{12} - \omega_d P_2.
\end{equation}
As $P_2$ oscillates with the frequency $\Omega$,
 {then the product $\omega_d P_2$ contains the
non-oscillating terms (see section 5.3)} and contribution to
$P_3$, which is caused by the EDM, linearly grows with time, when
$\Omega_f=\Omega$.
If $\Omega_f \ne \Omega$, then contribution to $P_3$ caused by EDM
slowly oscillates with the frequency $\Omega_f - \Omega$.

It is important that modulation of the velocity $v=v_0+\delta v
\sin{(\Omega_f t + \varphi_f)}$ results in $E_{eff}$ oscillation
(see (\ref{Eeff})) and, therefore, $E_{eff}^2$ also oscillates
with time and appears proportional to $\sin^2 {(\Omega_f t +
\varphi_f)}$.
 As a result, the contribution to $P_3$ caused by the tensor
 electric polarizability can be expressed as:
\begin{equation}
\frac{dP_3(\alpha_T)}{dt} \sim \Delta \Omega_T \sin^2 {(\Omega_f t
+ \varphi_f)}[ P_{12}(0)\cos{2 \Omega
t}-\frac{P_{11}(0)-P_{22}(0)}{2} \sin{2\Omega t} ],
\end{equation}
 where $\Delta \Omega_T = - \frac{2}{3} \frac{\alpha_T}{\hbar} (\delta E_{eff})^2$ i.e.
\begin{equation}
\frac{dP_3(\alpha_T)}{dt} \sim - \frac{1}{2} \Delta \Omega_T \cos
{(2 \Omega_f t+ 2\varphi_f)}[ P_{12}(0)\cos{2 \Omega
t}-\frac{P_{11}(0)-P_{22}(0)}{2} \sin{2\Omega t} ] \label{28}
\end{equation}

According to (\ref{28}) for a partially polarized deuteron beam
the derivative $\frac{dP_3}{dt}$
depends on the deuteron polarization components $P_{12}$ and
$\frac{P_{11}(0)-P_{22}}{2}$.


For simplicity let us consider a deuteron beam in the pure
polarization state.
In this case the components $P_{12}$ and
$\frac{P_{11}(0)-P_{22}}{2}$ can be written using the explicit
expression for the spin wavefunctions.
Suppose $\vec{n}(\vartheta, \varphi)$ is the unit vector directed
along the deuteron spin  ($\vartheta$ and $\varphi$ are the polar
and azimuth angles (see Fig.\ref{coordinate})).
  So the spin wavefunction that
describes the deuteron spin state with the magnetic quantum number
$m=1$
  can be expressed as follows (in the Cartesian
 basis):
\begin{eqnarray}
\chi_{1}(\vartheta, \varphi)=\left(
\begin{array}{c}
a_x\\
a_y\\
a_z
\end{array}
\right)=-\frac{1}{\sqrt{2}} \left(
\begin{array}{c}
\cos \vartheta \cos \varphi - i \sin \varphi\\
\cos \vartheta \sin \varphi + i \cos \varphi\\
- \sin \vartheta
\end{array}
\right)
\end{eqnarray}

Polarization vector can be written as:
\begin{equation}
\vec{P}=\langle \hat{S} \rangle=\chi_1^+ \hat{S} \chi_1 = i
[\vec{a} \times \vec{a}^{*}]
\end{equation}
and components of polarization tensor
\begin{equation}
\langle {P}_{ik}\rangle=\chi_1^+ \hat{Q}_{ik} \chi_1= -\frac{3}{2}
\{ a_i a_k^* + a_k a_i^* - \frac{2}{3}\},
\end{equation}
where $\hat{Q}_{ik}$ is the spin tensor of rank two
($\hat{Q}_{ik}=\frac{1}{2}(\hat{S}_i \hat{S}_k + \hat{S}_k
\hat{S}_i - \frac{4}{3} \delta_{ik} S)$).
Therefore,
\begin{eqnarray}
P_{12}=\frac{3}{4} \sin 2 \varphi \sin^2 \vartheta, \\
\frac{P_{11}-P_{22}}{2}=\frac{3}{4} \cos 2 \varphi \sin^2
\vartheta, \\
P_{33}=-\frac{3}{2} \left( \sin^2 \vartheta - \frac{2}{3}\right).
\end{eqnarray}

Using (49,53,54) one can obtain:
\begin{eqnarray}
\frac{dP_3 (\alpha_T) }{dt} \sim - \frac{3}{8} \Delta \Omega_T
\sin^2 \vartheta \cos (2 \Omega_f t + 2 \varphi_f) \times [\sin 2
\varphi \cos 2 \Omega t - \cos 2 \varphi \sin 2 \Omega t]
\label{101}
\end{eqnarray}
From (\ref{101}) it follows that $\frac{dP_3}{dt}$ slowly
oscillates with the frequency $(\Omega_f - \Omega)$.

In the ideal case, when $\Omega_f = \Omega$ (as it is proposed in
\cite{orlov} for EDM measurement) (\ref{101}) converts to:
\begin{eqnarray}
\frac{dP_3 (\alpha_T)}{dt} = - \frac{3}{8} \Delta \Omega_T \sin^2
\vartheta \cos (2 \Omega t + 2 \varphi_f) \sin ( 2 \Omega t -  2
\varphi) \label{102}
\end{eqnarray}
In the general case, when the phases $\varphi_f$ and $\varphi$ are
arbitrary, (\ref{102}) contains terms that do not depend on time
and, therefore, $P_3$ linearly grows with time, like the signal
from the EDM does.

It is interesting that making $\varphi_f=-\varphi$ provides:
\begin{eqnarray}
\frac{dP_3 (\alpha_T)}{dt}  \sim \cos (2 \Omega t - 2 \varphi)
\sin ( 2 \Omega t -  2 \varphi)=\frac{1}{2} \sin (4 \Omega t - 4
\varphi) \label{103}
\end{eqnarray}
 {that makes this contribution to $P_3$ quickly oscillating
and depressed}.
But even in this ideal case  it rests the contribution caused by
the tensor magnetic polarizability (\ref{VB}) (in real situation
$\Omega \ne \Omega_f$, though).

Measurement of these contribution provides to measure the tensor
electric polarizability.

According to evaluations \cite{polarizability} $\alpha_T \sim
10^{-40}$ cm$^3$, therefore, for the field $E_{eff}=B \sim 10^4$
gauss the frequency $ \Omega_T \sim 10^{-5}$ sec$^{-1}$.
When considering modulation one should estimate $\Delta \Omega_T
\sim \Omega_T {(\frac{\delta v}{v_0})}^2$, then  suppose
${(\frac{\delta v}{v_0})}^2 \sim 10^{-2} - 10^{-3}$ we obtain
$\Delta \Omega_T \sim 10^{-7} - 10^{-8}$ sec$^{-1}$,
 {that exceeds the magnitude of $\omega_d$ for the deuteron
EDM $d=10^{-29}~e \cdot cm$}.

\section{Particle spin rotation in a storage ring with an internal target}

Let now a gas target be placed inside the storage ring.
According to the section 5.1, spin evolution in this case is also
influenced by the nuclear pseudoelectric and pseudomagnetic
fields.

Time evolution of the spin and  tensor polarization can be found
by the equations:
\begin{equation}\label{vecP}
\vec{P} = \frac{\langle \Psi (t) | \frac{\vec{S}}{S} | \Psi (t)
\rangle}{\langle \Psi (t) | \Psi (t) \rangle};
P_{ik} = \frac{\langle \Psi (t) | \hat{Q}_{ik} | \Psi (t)
\rangle}{\langle \Psi (t) | \Psi (t) \rangle}
\end{equation}
$\hat{Q}_{ik}$ is the tensor of rank two (the tensor
polarization), for $S=1$ the tensor $\hat{Q}_{ik}= \frac{3}{2}(S_i
S_k + S_k S_i -\frac{4}{3} \delta_{ik} )$.

 But it should be taken
into account that multiple scattering, which occurs in the target,
brings to both change in the angular spread and depolarization of
the particle beam.

So the spin density matrix formalism should be used to derive
equations for evolution of the particle spin.

The density matrix of the system "deuteron+target" is
\begin{eqnarray}
\rho=\rho_{p}\otimes\rho_{t},
 \label{2.6}
\end{eqnarray}
where $\rho_{d}$ is the density matrix of the particle beam and
$\rho_{t}$ is the density matrix of the target.

The equation for the particle density matrix can be written as
\cite{4rot,nastya}:
\begin{eqnarray}
\frac{d\rho_{p}}{dt}=-\frac{i}{\hbar}\left[\hat{H}_p,\rho_{p}\right]+\left(\frac{\partial\rho_{p}}{\partial
t}\right)_{col},
\label{2.9}
\end{eqnarray}
where $\hat{H}_p$ is the Hamiltonian describing particle motion in
external electromagnetic fields,
$\left(\frac{\partial\rho_{p}}{\partial t}\right)_{col}$ describes
evolution of the density matrix due to collisions with the target
atoms (nuclei).

 The collision term
$\left(\frac{\partial\rho_{d}}{\partial t}\right)_{col}$ can be
expressed as follows:
\begin{eqnarray}
\left(\frac{\partial\rho_{d}}{\partial t}\right)_{col} =v N
\textrm{{Sp}}_{t}\left[\frac{2\pi i}{k}\left[F(\theta=0)\rho-\rho
F^{+}(\theta=0)\right] +\int d\Omega
F(\vec{k}^{'})\rho(\vec{k}^{'})F^{+}(\vec{k}^{'})\right],
\label{2.11}
\end{eqnarray}
where $\vec{k}^{'}=\vec{k}+\vec{q}$, {$\vec{q}$ is the momentum
transferred to a nucleus of matter from the incident particle,}
 $v$ is the speed of the incident particles, $N$ is the
atom density in matter,
 $F$ is the scattering amplitude
depending on the spin operators of the deuteron and the matter
nucleus (atom), $F^+$ is the Hermitian conjugate for the operator
$F$.
{The first term in (\ref{2.11}) describes coherent elastic
scattering of a particle by matter nuclei, while the second term
is for multiple scattering.}
 Let us consider the first term in (\ref{2.11}). It can be written
as:
\begin{eqnarray}
\left(\frac{\partial\rho_{p}}{\partial t}\right)_{col}^{(1)}=v N
\frac{2\pi i}{k}
 \left[
 \hat{f}(0)\rho_p-\rho_p \hat{f}(0)^{+}
\right] .
\label{2.12}
\end{eqnarray}
The  amplitude $\hat{f}(0)$ of particle scattering in the target
at the zero angle is
\begin{eqnarray}
\hat{f}(0)=\textrm{{Sp}}_{t}\hat{F}(0)\rho_{t}.
\label{2.13}
\end{eqnarray}

As a result one can obtain:
\begin{eqnarray}
\left(\frac{\partial\rho_{p}}{\partial t}\right)_{col}^{(1)} =
 -\frac
i\hbar\left(\hat{V}_{eff} {\rho_p}-{\rho_p}
\hat{V}_{eff}^{+}\right).
\label{2.15}
\end{eqnarray}
where
\begin{equation}
\hat{V}_{eff}=-\frac{2 \pi \hbar^2}{m \gamma} N
\hat{f}(0)=-\frac{2 \pi \hbar^2}{m \gamma} N (d +A_1 \vec{S}
\vec{P}_t + A_2 (\vec{S} \vec{n})(\vec{n} \vec{P}_t) + d_1
(\vec{S} \vec{n})^2). \label{82}
\end{equation}
The latter expression can be rewritten as:
\begin{equation}
\hat{V}_{eff}=- \frac{2 \pi \hbar^2}{m \gamma} N d - \frac{\mu}{S}
\vec{S} \vec{G} - \frac{2 \pi \hbar^2}{m \gamma} N d_1 (\vec{S}
\vec{n})^2) . \label{83}
\end{equation}
where $\vec{G}=\frac{2 \pi \hbar^2 S}{m \gamma \mu} N (A_1
\vec{P}_t + A_2 \vec{n}(\vec{n} \vec{P}_t))$ (see (\ref{26})).
For particles with the spin $\frac{1}{2}$ the term proportional to
$d_1$ is absent.

Finally, the expression (\ref{2.9}) reads:
\begin{eqnarray}
\frac{d\rho_{p}}{dt}=-\frac{i}{\hbar}\left[\hat{H}_p,\rho_{d}\right]
 -\frac
i\hbar\left(\hat{V_{eff}} {\rho_p}-{\rho_p}
\hat{V}^{+}_{eff}\right)+
 v N
\textrm{{Sp}}_{t} \int d\Omega
F(\vec{k}^{'})\rho(\vec{k}^{'})F^{+}(\vec{k}^{'}).
 \label{2.9_new}
\end{eqnarray}
The last term in the above formula, which is proportional to
$\emph{Sp}_{t}$, describes the multiple scattering process and
spin depolarization aroused from it.

When studying interaction of particles with a target inside a
storage ring the density of the target is chosen to make multiple
scattering in it small for the observation time; depolarization of
the beam also appears small.

For high-energy particles scattering at small angles is important.
Therefore the large number of scattered particles remain on their
orbits. For this case the latter term should be considered.

Further analysis of equations (\ref{2.9_new}) in the present paper
will be done for such time of the experiment that provides the
last term in (\ref{2.9_new}) can be cast out.

As a result we have:
\begin{eqnarray}
\frac{d\rho_{p}}{dt}=-\frac{i}{\hbar}\left[\hat{H}_p,\rho_{d}\right]
 -\frac
i\hbar\left(\hat{V_{eff}} {\rho_p}-{\rho_p}
\hat{V}^{+}_{eff}\right). \label{rho0}
 \end{eqnarray}

Now the equation for evolution of spin properties of a particle in
a storage ring can be obtained.
The polarization vector is as follows:
\begin{equation}
\vec{P}=\frac{\textrm{Sp}~\rho_p (t)
~\frac{\vec{S}}{S}}{\textrm{Sp}~\rho_p
(t)}=\frac{\textrm{Sp}~\rho_p (t) ~\vec{S}}{I(t)~{S}},
\label{vecPnew}
\end{equation}
where $I(t)=\textrm{Sp}~\rho_p (t)$ is the beam intensity.

From (\ref{vecPnew}) one can get the differential equation
providing to find the beam polarization:

\begin{equation}
\frac{d \vec{P}}{dt}=\frac{\textrm{Sp}~  \frac{d \rho_p (t)}{dt}
~{\vec{S}}}{{S} I(t)} -  \vec{P} \frac{1}{I(t)} ~ \frac{d
I(t)}{dt} \label{dP}
\end{equation}

When the particle spin is $S=1$ then the polarization tensor
should also be found:
\begin{equation}
P_{ik}=  \frac{\textrm{Sp}~   \rho_p ~\hat{Q_{ik}}}{I(t)},
\label{Piknew}
\end{equation}

Change in the tensor polarization with time can be written as:
\begin{equation}
\frac{d P_{ik}}{dt}=\frac{1}{I(t)} \textrm{Sp}\left( \frac{d
\rho_p}{dt} \hat{Q_{ik}} \right) - P_{ik} \frac{1}{I(t)} ~\frac{d
I(t)}{dt}. \label{dPik}
\end{equation}

Using the equation for the density matrix (\ref{2.9_new}) and the
expression (\ref{dP},\ref{dPik}) one can obtain the equation that
summarizes BMT equation (\ref{BMT}) for the case when spin
evolution is influenced by the pseudoelectric and pseudomagnetic
nuclear fields.

Let us consider first particles with the spin $S=\frac{1}{2}$.
The density matrix for such a particle can be expressed as
follows:
\begin{equation}
\rho_{p \frac{1}{2}}=I_{\frac{1}{2}}(t) (\frac{1}{2} \hat{I} +
\vec{P} \vec{S}),\label{rho12}
\end{equation}
where $\hat{I}$ is the unit matrix in the spin space.

With the help of (\ref{vecPnew}) it is possible to find that
\begin{equation}
\frac{d I_{\frac{1}{2}}(t)}{dt}=-\frac{i}{\hbar} \textrm{Sp}
(\hat{V}_{eff}\rho_{p \frac{1}{2}}-\rho_{p
\frac{1}{2}}\hat{V}_{eff}^+)=-(\varkappa + \frac{2 \mu}{\hbar}
\vec{G}_2 \vec{P}) I_{\frac{1}{2}}(t), \label{dI}
\end{equation}
where $\varkappa= v N \sigma_{tot}$, $v$ is the particle speed,
$\sigma_{tot}$ is the total cross-section of particle scattering
by a nonpolarized nucleus and $\vec{G}_2$ is the imaginary part of
the pseudomagnetic nuclear field.
The expression (\ref{dI}) can be rewritten as:
\begin{equation}
\frac{d I_{\frac{1}{2}}(t)}{dt}=-\varkappa_{abs} (\vec{P})
I_{\frac{1}{2}}(t), \label{dI1}
\end{equation}
where the absorption coefficient $\varkappa_{abs}=v N \left[
\sigma_{tot}+  \frac{1}{2}\sigma_{1} (\vec{P}\vec{P}_t) +
\frac{1}{2}\sigma_{2} (\vec{P}\vec{n})(\vec{n}\vec{P}_t)\right]=
\varkappa + \vec{P}\vec{g}_t$, $\vec{g}_t=\frac{1}{2}v N (
\sigma_{1} \vec{P}_t + \sigma_{2}\vec{n})(\vec{n}\vec{P}_t)$.

The cross-section $\sigma_{1}=\sigma_{tot}^{\uparrow \uparrow}
(\vec{n} \perp \vec{P}_t) - \sigma_{tot}^{\uparrow \downarrow}
(\vec{n} \perp \vec{P}_t)$ is the difference between the
cross-sections of particle scattering by a polarized nucleus with
the $\vec{P}$ parallel ($\vec{P} \uparrow \uparrow \vec{P}_t$) and
antiparallel ($\vec{P} \uparrow \downarrow \vec{P}_t$) to the
target polarization vector $\vec{P}_t$ in conditions when the
particle momentum is orthogonal to the target polarization vector
$(\vec{n} \perp \vec{P}_t)$.
The cross-section $\sigma_{2}=\sigma_{tot}^{\uparrow \uparrow}
(\vec{n} || \vec{P}_t) - \sigma_{tot}^{\uparrow \downarrow}
(\vec{n} || \vec{P}_t)$ is the difference between the
cross-sections of particle scattering by a polarized nucleus with
the $\vec{P}$ parallel ($\vec{P} \uparrow \uparrow \vec{P}_t$) and
antiparallel ($\vec{P} \uparrow \downarrow \vec{P}_t$) to the
target polarization vector $\vec{P}_t$ in conditions when the
particle momentum is parallel to the target polarization vector
$(\vec{n} || \vec{P}_t)$.

The equation (\ref{dI1}) describes the well known fact that the
coefficient of particle beam absorption in a polarized target
depends on the respective orientations of the beam and target
polarization vectors.

Let us obtain now the equation that describes evolution of the
particle vector polarization under action of pseudomagnetic
nuclear field $\vec{G}$.

According to (\ref{vecPnew},\ref{dP}) one can obtain:
\begin{equation}
\frac{d \vec{P}}{dt}= \frac{1}{S I_{\frac{1}{2}}(t)} \textrm{Sp}
\frac{d \rho_p}{dt} \vec{S} - \vec{P} \frac{1}{I_{\frac{1}{2}}(t)}
\frac{dI_{\frac{1}{2}}(t)}{dt} = -\frac{2 \mu}{ \hbar} \left[
\vec{G}_1 \times \vec{P} \right] - \frac{2 \mu}{\hbar} (\vec{G}_2
- \vec{P} (\vec{G}_2 \vec{P})), \label{dP2}
\end{equation}
where $vec{G}_1= \textrm{Re} \vec{G}$, $\vec{G}_2= \textrm{Im}
\vec{G}$.

The equation (\ref{dP2}) can be rewritten as:
\begin{equation}
\frac{d \vec{P}}{dt} = \left[ \vec{P} \times \vec{\Omega}_{nuc}
\right]- (\vec{g}_t - \vec{P} (\vec{P} \vec{g}_t)), \label{dP3}
\end{equation}
where $\vec{\Omega}_{nuc}$ is the frequency of spin precession in
the pseudomagnetic nuclear field
\begin{equation}
\vec{\Omega}_{nuc}= \frac{2 \mu}{ \hbar}  \vec{G}_1 = \frac{2 \pi
\hbar}{ m \gamma} N (\textrm{Re} A_1 \vec{P}_t + \textrm{Re} A_2
\vec{n} ( \vec{n} \vec{P}_t)). \label{Onuc}
\end{equation}
Adding the contribution from the pseudomagnetic nuclear field
(\ref{dP3}) to the BMT equation (\ref{BMT}) one can finally obtain
the equation describing spin evolution for particles moving in a
storage ring with a polarized target inside:
\begin{eqnarray}
\frac{d \vec{P}}{dt} = \left[ \vec{P} \times (\vec{\Omega}(d)+
\vec{\Omega}_{nuc}) \right]- (\vec{g}_t - \vec{P} (\vec{P}
\vec{g}_t)), \\
\frac{d I_{\frac{1}{2}}(t)}{dt}=-(\varkappa + \vec{P} \vec{g}_t )
I_{\frac{1}{2}}(t),
 \label{dP4}
\end{eqnarray}
where $\vec{g}_t =\frac{1}{2} v N (\sigma_1 \vec{P}_t + \sigma_2
\vec{n} ( \vec{n} \vec{P}_t))$.

The contribution from the pseudomagnetic nuclear field to spin
evolution of the $S=1$ particle (for example, deuteron) can be
obtained in the similar way. The spin density matrix for the spin
1 particle is expressed as follows:
\begin{equation}
\rho_1=I_1(t)(\frac{1}{3} \hat{I} + \frac{1}{2} (\vec{P} \vec{S})
+ \frac{1}{9} P_{ik} \hat{Q}_{ik})
 \label{rho1}
\end{equation}

The effective potential energy of deuteron interaction with a
polarized target is expressed as $\hat{V}_{eff}=- \frac{2 \pi
\hbar^2}{m \gamma} N d - \frac{\mu}{S} \vec{S} \vec{G} - \frac{2
\pi \hbar^2}{m \gamma} N d_1 (\vec{S} \vec{n})^2$ $~~$(see
(\ref{83})).

The energy $\hat{V}_{eff}$ for a deuteron (in contrast to a
particle with the spin $S=\frac{1}{2}$ ) contains both deuteron
interaction with the pseudomagnetic field and pseudoelectric field
(the term proportional to $d_1$).
The equations describing deuteron spin rotation (birefringence
effect) in the pseudoelectric field of the nonpolarized internal
target of a storage ring were obtained in \cite{nastya}.

If the internal target in a storage ring is polarized, then using
(\ref{83},\ref{rho0},\ref{rho1}) one can obtain the change in the
beam intensity and polarization vector as follows:
\begin{equation}
\frac{dI_1(t)}{dt}=-\left[ (\varkappa + 2 \vec{g}_t
\vec{P})-\frac{\chi}{3}(2+P_{ik} n_i n_k) \right] I_1 (t)
\label{dI11}
\end{equation}
where $\chi=-\frac{4 \pi v N}{k} \textrm{Im} d_1=-v N
(\sigma_{M=1}-\sigma_{M=0})$, $\varkappa=v N \sigma_{M=0}$,
$\sigma_{M=1}$ and $\sigma_{M=0}$ are the total cross-sections of
deuteron scattering by a nonpolarized nucleus for the deuteron
state with the magnetic quantum number $M=1$ and $M=0$,
respectively (the quantization axis is directed along $\vec{n}$).


The polarization vector can be expressed as:
\begin{eqnarray}
\begin{array}{c}
\frac{d \vec{P}}{dt}= \left[ \vec{P} \times
\vec{\Omega}_{nuc}\right] - \frac{2}{3}  P_{ik} {g_t}_k -
\frac{4}{3} \vec{g}_t + 2 \vec{P} (\vec{P} \vec{g}_t) +
\frac{\eta}{3}[\vec{n}\times\vec{n}^{'}]+ \\
+\frac{\chi}{2}(\vec{n}(\vec{n}\cdot\vec{P})+\vec{P})
-\frac{2\chi}{3}\vec{P}-\frac{\chi}{3}(\vec{n}\cdot\vec{n}^{'})\vec{P}
\end{array}
\label{dP11}
\end{eqnarray}
where $\eta=-\frac{4 \pi N}{k} \textrm{Re }d_{1}$,
$n_{i}^{'}=P_{ik}n_{k}$.


According to (\ref{dP11}) the nuclear pseudomagnetic field causes
deuteron spin precession with the frequency $\vec{\Omega}_{nuc}$.

Contribution (\ref{dP11}) should be added to the equation
(\ref{BMT+}) to make them describing evolution of the deuteron
spin in a storage ring at presence of target.

The similar approach provides to find the addition to the tensor
polarization evolution $\frac{d P_{ik}}{dt}$ (see expression
(\ref{dPik})).
But it is too bulky to adduce it here.

However, when particle beam absorption in the target can be
neglected, one can make all changes in the equation, which
describes deuteron spin behavior, by replacement of
$\vec{\Omega}(d)$ with new value
$\vec{\Omega}(d)+\vec{\Omega}_{nuc}$.

Note that as the target inside the storage ring is of the finite
size then the particle moving in the storage ring interacts with
the target {at times} (the expressions for $\vec{\Omega}_{nuc}$
and $\vec{g}_t$ contain the density $N$, which means the density
in the point of particle location i.e. $N=N(t)$ ).
This is the reason for $\vec{\Omega}_{nuc}$ and $\vec{g}_t$
oscillating with the frequency of particle rotation in the storage
ring $\Omega_0$.

For further analysis let us expand  $\vec{\Omega}_{nuc}$ and
$\vec{g}_t$ into Fourier series and {be confined }with the zero
harmonics. In this case $\vec{\Omega}_{nuc}$ and $\vec{g}_t$
appears to be constants and can be written as:
\begin{eqnarray}
\begin{array}{l}
\vec{\Omega}_{nuc}=
 \frac{2 \pi \hbar}{ m \gamma} N
(\textrm{Re} A_1 \vec{P}_t + \textrm{Re} A_2 \vec{n} ( \vec{n}
\vec{P}_t)) \frac{l}{vT}
=\\
=\frac{2 \pi \hbar}{ m \gamma} \frac{j_t }{L} (\textrm{Re} A_1
\vec{P}_t + \textrm{Re} A_2 \vec{n} ( \vec{n} \vec{P}_t))
= \\
=\frac{2 \pi \hbar}{ m \gamma} \frac{j_t \nu}{v} (\textrm{Re} A_1
\vec{P}_t + \textrm{Re} A_2 \vec{n} ( \vec{n} \vec{P}_t)),
\end{array}
\end{eqnarray}
where $l$ is the target length, $j_t=N \cdot l$ is the target
density in cm$^{-2}$ (usually for a polarized gas target $j_t
\approx 10^{14}$ cm$^{-2}$), $T$ is the period of beam rotation in
the storage ring, $\nu$ is the frequency of particle rotation in
the storage ring, $L$ is the particle orbit length inside the
storage ring.

The zero harmonics for $\vec{g}_t$ is as follows:
\begin{eqnarray}
\begin{array}{l}
\vec{g}_t=
 \frac{1}{2} v N
(\sigma_1 \vec{P}_t + \sigma_2 \vec{n} ( \vec{n}
\vec{P}_t)) \frac{l}{vT} = \\
=\frac{1}{2} {j_t \nu} (\sigma_1 \vec{P}_t + \sigma_2 \vec{n} (
\vec{n} \vec{P}_t)).
\end{array}
\end{eqnarray}

Let us evaluate now the effect magnitude. For protons, antiprotons
and deuterons with the energy from MeV to GeV $\textrm{Re} A_1
\sim 10^{-12} \div 10^{-13}$ cm.
Considering $\nu \sim 10^6$ s$^{-1}$ one gets for the nuclear
precession frequency $\Omega_{nuc}=10^{-4} \div 10^{-5}$ s$^{-1}$.

Note (see (\ref{Veff_new},\ref{Veff_new1})) that the ratio
$\frac{1}{\gamma} \textrm{Re} A_{1(2)}$ is proportional to the
T-matrix and, as it follows, $\Omega_{nuc}$ depends on the
particle energy only due to possible dependence of the T-matrix on
energy.

The obtained estimation for $\Omega_{nuc}$ allows to expect, for
example for COSY, to observe the spin rotation angle $\vartheta =
\Omega_{nuc} t \approx 10^{-1} \div 10^{-2}$ rad in the
observation time $t \sim 10^3$ s. This value is quite observable.

To prevent suppression of the the spin precession in the
pseudomagnetic nuclear field by the storage ring magnetic field
$\vec{B}$ the polarized target should be placed in the straight
section of the storage ring, where the field $\vec{B}=0$ and the
particle moves along the straight trajectory.

Let us now consider the particular example.
Suppose the axis $OZ$ is orthogonal to the orbit plane ($OS ||
\vec{B}$) (see Fig.2).
The pseudomagnetic field is directed along the axis $OX$.
In this case for the particle in the straight section the storage
ring the vertical spin component rotates around the direction of
the pseudomagnetic field in the $ZOY$ plane.
Just the change of the vertical spin component should be observed.

As the typical spin rotation angle $\vartheta = \Omega_{nuc} t \ll
1$, then change of the vertical spin component with time can be
expressed as:
\begin{equation}
P_3(t)=P_3(0) - \frac{1}{2} \vartheta^2 t^2 = P_3(0) -
\Omega_{nuc}^2 t^2. \label{P3comp}
\end{equation}

The effect can be strengthened when adding a magnetic field
$\vec{b} \gg \vec{G}_1$ directed along the pseudomagnetic field
$\vec{G}_1$ (along the axis $OX$ in the case under consideration).
In this case, the angle of rotation is expressed as $\vartheta =
\vartheta_{mag} + \vartheta_{nuc} = ( \Omega_{mag} (\vec{b}) +
\Omega_{nuc}) t$ and the vertical component $P_3(t)=P_3(0) -
\frac{1}{2} \vartheta_{mag}^2 t^2 - \vartheta_{mag}
\vartheta_{nuc}$ (the terms $\sim \vartheta_{nuc}^2$ are neglected
comparing with the previous terms).

When make the direction of the field $\vec{b}$ changing during the
experiment (i.e. the sign of $\vartheta_{mag}$ changing) one gets
the possibility to measure the difference $P_3 (\vec{b} \uparrow
\uparrow \vec{G}_1,t) - P_3 (\vec{b} \uparrow \downarrow
\vec{G}_1,t) = 2 \vartheta_{mag} \vartheta_{nuc}$.
This provides to study the possibility of effect observation and
measurement of $\vartheta_{nuc}$ and, therefore, to measure the
spin-dependent part of the amplitude of elastic coherent
zero-angle scattering.

\section{Conclusion}

In the present paper the equations for spin evolution of a
particle in a storage ring are obtained considering contributions
from the tensor electric and magnetic polarizabilities of the
particle along with the contributions from spin rotation and
birefringence effect in polarized matter of a gas target.

 Influence of the tensor electric and
 magnetic polarizabilities on spin evolution in the resonance deuteron EDM
 experiment  is considered in details.

 It is shown that, besides the EDM, the electric and magnetic polarizabilities also contribute
 to the vertical spin component $P_3$. Moreover, the electric polarizability contributes
 to the $P_3$ component even when the deuteron EDM is supposed to be zero and thereby
 the electric polarizability can imitate the EDM contribution.
 It is shown that unlike the vertical component of the spin $P_3$
 the component $P_{33}$ of polarization tensor does not contain
 contribution from the electric polarizability, whereas contribution from the magnetic
 polarizability reveals only when the deuteron EDM differs from zero.

It is also shown that when the angle $\vartheta$ between the spin
direction and the vertical axis meets the condition $\sin
\vartheta=\sqrt{\frac{2}{3}}$ ($\cos
\vartheta=\sqrt{\frac{1}{3}}$), the initial value of $P_{33}$
appears $P_{33}(0)=0$.
As a result, the EDM contribution to the measured signal linearly
growth in time starting from zero that is important for
measurements.

Therefore, measurement of the $P_{33}$ component of the deuteron
tensor polarization seems to be of particular interest, especially
because  appearance of the nonzero component $P_{33}$  on its own
indicates the EDM presence (in contrast to the $P_3$ component,
which appearance can be aroused by the tensor electric
polarizability, rather than EDM).

It is also shown that study of spin rotation and the birefringence
effect for a particle in a high energy storage ring provides for
measurement of the spin-dependent real part of the coherent
elastic zero-angle scattering amplitude.


\end{document}